\documentclass{aastex}

\usepackage{graphicx}

\usepackage{emulateapj5}
\usepackage{times}

\makeatletter

\newenvironment{inlinefigure}{%
\def\@captype{figure}%
\noindent\begin{minipage}{0.999\linewidth}\begin{center}}
{\end{center}\end{minipage}\smallskip}
\makeatother

\def\keV{ke\kern-0.05emV}
\newcommand{\chandra}{\emph{Chandra}}

\renewcommand{\arcsec}{\ensuremath{''}}

\def\asca       {{\em ASCA}\/}
\def\chandra    {{\em Chandra}\/}
\def\rxj1720    {{ RXJ1720.1+2638}\/}

\def\rosat      {{\em ROSAT}\/}

\def\degd       {$^{\circ}\!$}
\def\second     {{\prime\prime}}
\def\ms1455     {{MS 1455.0+2232}}

\begin{document}

%\submitted{Submitted to ApJ Letters}

\title{\chandra ~ Observation of \ms1455 ~: 
cold fronts in a massive cooling flow cluster?}

\author{P.\ Mazzotta\altaffilmark{1,2}, M. Markevitch \altaffilmark{1},
 W.R. Forman\altaffilmark{1}, C. Jones\altaffilmark{1}, 
 A. Vikhlinin \altaffilmark{1,3}, 
and L. VanSpeybroeck \altaffilmark{1}}
\altaffiltext{1}{Harvard-Smithsonian Center for Astrophysics, 60 Garden St.,
Cambridge, MA 02138; mazzotta@cfa.harvard.edu}
\altaffiltext{2}{ESA Fellow}
\altaffiltext{3}{Space Research Institute, Russian Academy of Science}

%\shorttitle{}
\shortauthors{MAZZOTTA ET AL.}

\begin{abstract}

We present the \chandra ~ observation of the cluster of galaxies
\ms1455 . From previous \asca ~ and \rosat ~ observations, 
this cluster was identified as a 
``relaxed'' cluster that hosts
one of the most massive cooling flows detected. We observe a
sharp brightness peak and a temperature decrease toward the center.
With higher angular resolution,
the \chandra ~ X-ray image shows the presence of two surface brightness
edges on opposite sides of the X-ray peak: the first, with  
a surface brightness jump of a factor $\approx 10$, 
at $190~h_{50}^{-1}$~kpc to the north
and the second, with a jump factor of $\approx 3$, at $450~h_{50}^{-1}$~kpc
to the south. 
%The observed edges are 
 Even though the low exposure of
this observation
limits our ability to constrain the temperature jump across both edges,
we show that the northern edge is likely to be a ``cold front'' similar to
others observed recently by \chandra ~ in the clusters
A2142, A3667, RX J1720.1+2638, and A2256. The observed 
cold front is most likely
produced by the motion, from  south 
to  north, of a group-size dark matter halo.
The most natural explanation for the presence of this observed moving 
subclump is that \ms1455 ~ is a merger cluster in the very last stage 
before it becomes fully relaxed. 
This scenario, however, appears to be unlikely as 
the cluster shows no further sign of ongoing merger. Moreover, 
it is  not clear if a massive cooling flow could have survived 
this kind of merger.  
We propose an alternative scenario
 in which,  as for 
RX J1720.1+2638,  \ms1455 ~ is  the result 
of the hierarchical collapse 
of two co-located density perturbations, the first
 a group-scale perturbation collapse followed by a second 
cluster-scale perturbation collapse
that surrounded, but did not destroy, the first.
We suggest that a cooling flow may have begun inside
the already collapsed group-scale perturbation and may have been later
amplified by the gas compression induced by the infall of the 
overlying main cluster mass.

\end{abstract}


\section{Introduction}

With its high angular resolution, \chandra ~ has recently 
discovered surface brightness edges in several clusters of galaxies. 
Even though surface brightness edges are theoretically expected in 
shock fronts, these edges are indeed the result of 
a new unpredicted phenomenon.
In fact,  they  show  a temperature variation across the edge that 
goes exactly in the opposite direction to what is predicted in a shock front:
the temperature in  front of the shock (on the less bright side of the
edge) is higher than the  ``post shock''
temperature (on the brighter region of the edge). 
For this reason these newly discovered edges have been called ``cold
fronts'' (Vikhlinin, Markevitch, \& Murray 2000a).
As the first cold fronts were discovered in two merging clusters A2142
(Markevitch et al. 2000b) and A3667 (Vikhlinin et al. 2000a,b), it has been 
suggested that  they  are produced by the relative motion of the gas cloud
of the merging subclump with respect to the gas of the main clusters. 
Successively  \chandra ~  revealed the presence of a cold front in
the cluster of galaxies RX J1720.1+2638. Unlike the first two, the
X-ray image of RX J1720.1+2638 is  azimuthally symmetric at large radii,  
it shows no other signs of an
 ongoing merger, and the moving  gas cloud appears
to be centered on the cluster center (Mazzotta et al. 2001).
Moreover the  gas cloud speed is subsonic, much smaller than
that expected for a point mass free-falling from
infinity into the cluster center ($v\approx 2.7\times c_s$, where $c_s$ is the
speed of sound; see e.g. Sarazin 1988).   

To reconcile this apparent ``relaxed'' appearance of 
RX J1720.1+2638 with the presence of a moving gas cloud, it has been
proposed that either the cluster is in the last stage of a merger before 
it becomes fully relaxed or it is the result of the collapse of a
group of galaxies  followed by the 
collapse of a much larger, cluster-scale  perturbation at nearly the same
 location in space (Mazzotta et al. 2001). 

Both scenarios  raise a number of questions about the formation 
history of the cluster and/or the ``inefficiency'' of 
tidal disruption forces.  
To differentiate among the  possibilities and to better
understand the dynamics of  cluster formation, it is important to
search for evidence of motion, such as cold fronts, also in other apparently   ``relaxed'' 
clusters whose characteristics are similar to those of RX J1720.1+2638.

In this paper we present the \chandra ~ observation of
the cluster of galaxies \ms1455 . 
This cluster was previously observed  by \asca ~ and \rosat . 
From these observations it was classified as a 
``relaxed'' cluster hosting one of the
most  massive cooling flows observed  
($\dot M \approx 1500 M_\odot$~yr; Allen et al. 1996). 

The \chandra ~ image 
 reveals the presence of two surface brightness
edges  on opposite sides of the X-ray peak.
We show that one edge cannot be produced by a shock front but
rather indicates that it is a cold front.    
We discuss the possible implication of the presence of a cold front 
in a massive cooling flow cluster.

The structure of the paper is as follows. In $\S$~\ref{par:data} we 
describe the imaging and spectral analysis 
($\S$~\ref{par:image} and  $\S$~\ref{par:spectral}, respectively). 
In $\S$~\ref{par:nature}
 we discuss the nature of the
observed surface brightness edges, while in $\S$~\ref{par:dynamic} and
$\S$~\ref{par:mass} we discuss the
implications of the presence of a cold front for the cluster dynamics
and the X-ray mass determination, respectively. Finally in
$\S$~\ref{par:conclusion} we give our conclusions. 

We use $H_0=50$~km~s$^{-1}$~kpc$^{-1}$, $\Omega=1$, and $\Lambda=0$,
which imply a linear scale of
5.0~kpc per arcsec at the distance of MS1455.0+2232 ($z=0.258$).
Unless specified differently, all the errors are at $90\%$ confidence level
 for one interesting parameter.
%
% 
%

%\begin{figure*}
\begin{inlinefigure}
\centerline{\includegraphics[width=0.95\linewidth]{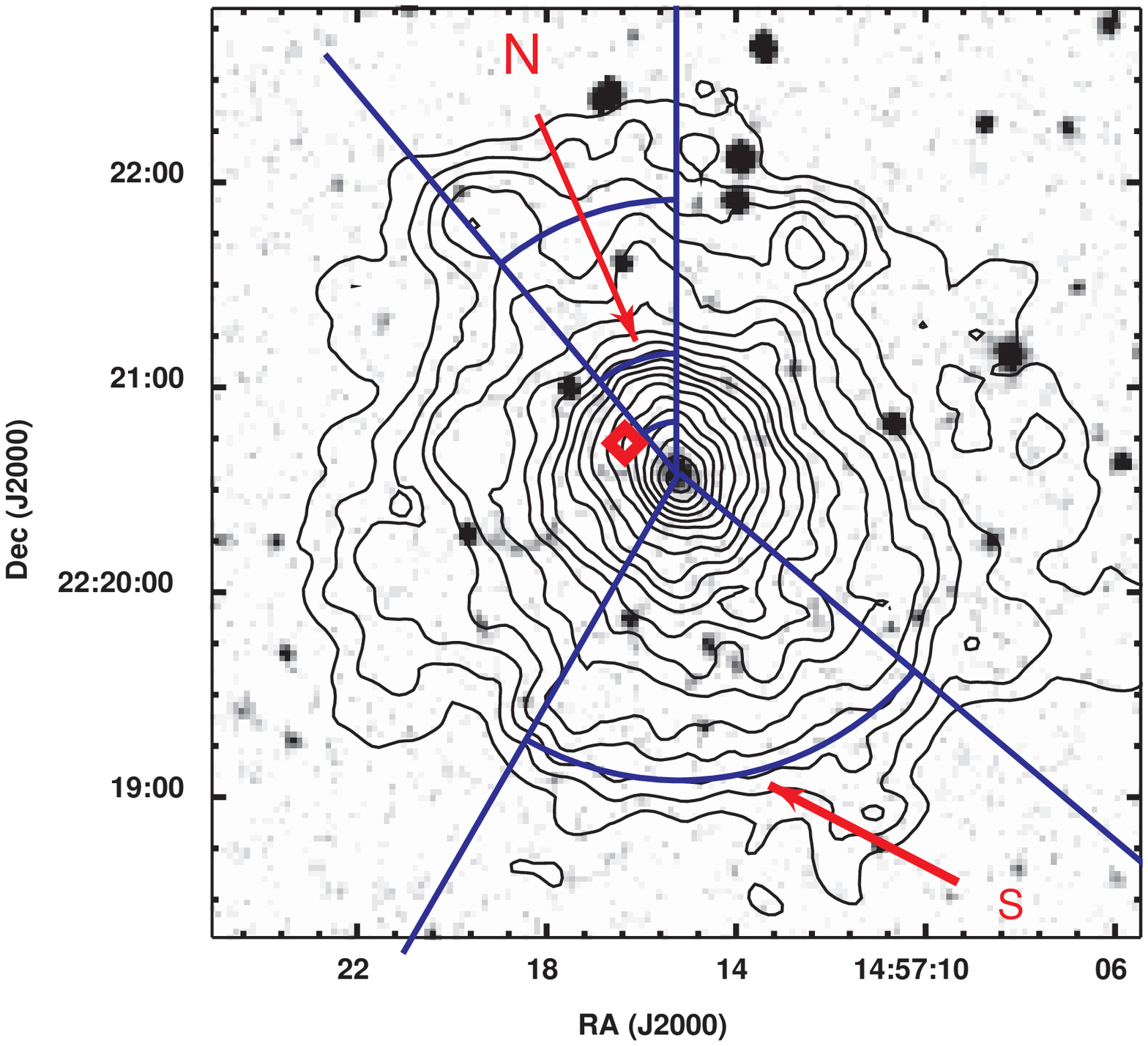}}
%\plotone{f1.eps}
\caption{Digitized Sky Survey image with overlaid  ACIS-I X-ray
surface brightness contours (log-spaced by a factor of $\sqrt{2}$) in the
0.5-8~keV energy band after adaptive smoothing. The four straight lines
starting from the X-ray brightness peak  define  
the northern and the southern  sectors whose angles are 0\degd~ to 40\degd ~ and
 150\degd~ to 230\degd  respectively
(the position angles are measured
from North toward East). The three red arcs in the north sector are at
located at $r=15\arcsec$,$35\arcsec$, and $80\arcsec$, respectively.
 The arrows indicate the northern and southern
surface brightness edges. The red diamond indicates the position of
the lensed galaxy. 
The central galaxy is 
coincident with the X-ray peak.}
\label{fig:optical}
%\end{figure*}
\end{inlinefigure}

\section{Data analysis}\label{par:data} 

\ms1455 ~ was observed  on May 2000 in  ACIS-I with an exposure of
$\approx 10$~ksec.
Hot pixels, bad columns, chip node boundaries, and events with
grades 1, 5, and 7 were excluded from the analysis. 
We extracted the  light curve from  chip S2, and 
verified that the observation presents no  strong background rate 
variations on time scales of a few hundred seconds.  
 
In our spectral and imaging analysis,
we used the public background dataset  composed of several 
observations of relatively empty fields taken from Feb. to Sept. 2000. These
observations were screened in exactly the same manner as the cluster data.
The total exposure of the background dataset is about 400ks.
The background spectra and images were  normalized by the ratio of the
respective exposures. 
%Following the prescription of Markevitch et al. (2000a) we also increased 
%the  background dataset by $5\%$ to take into account the observed
%reduction of the bekground {\bf (aggiusta)} 
This procedure yields a background that is accurate
to $\sim 10$\% based on comparison to other fields; this uncertainty is
taken into account in our analysis (see Markevitch et al. 2000a for a 
description of the ACIS background modeling).
%In any case we are interested in the very central region of the
%cluster ($r<150^\second$) and at these radii the contribution of the
%background to the total emission in the $0.5-8$~keV energy band is $<4\%$. 

%
%
%
\subsection{Imaging Analysis}\label{par:image}
To study the X-ray morphology of the cluster we generated 
an image with $1^\second \times 1^\second$ pixels 
from the events in the chips I0, I1, I2, and I3.
We extracted the image
in the 0.5-8~keV band and corrected for vignetting.  
In  Fig.~\ref{fig:optical} we  show the ACIS-I, X-ray
contours  
 (after an adaptive smoothing of the image with a circular top hat
filter with  a minimum of 20 counts under the
filter) on the DSS
optical image.
The X-ray brightness peak is located at ($14^{\rm h}57^{\rm m}15^{\rm
s}, \ \ +22^{\rm o}20^{\rm \prime}32.6^{\rm
\prime\prime} $). Thus it coincides, within 1~arcsec, 
with the optical center of the cluster central galaxy 
($14^{\rm h}57^{\rm m}15.1^{\rm
s}, \ \ +22^{\rm o}20^{\rm \prime}31^{\rm
\prime\prime}) $ (Allen et al. 1992).

The isointensity contours show that there is a steep 
gradient (edge) in the surface brightness at a distance of 
$r=38^\second$ ($190~h_{50}^{-1}$~kpc) from the center to the north.
To better visualize the surface brightness edge,
in Fig.~\ref{fig:photon} we show the photon image of the cluster in the 
$0.5-2.5$~keV energy band, binned to $2\arcsec$, after a Gaussian
smoothing with $\sigma=1$~pixel.
The image clearly shows that the edge remains sharp in a wide
 sector  whose angles are from 
$\approx  0$\degd~ to $\approx  40$\degd  ~ (hereafter northern-sector; 
the position angles are measured
from North toward East) before 
gradually vanishing.
We notice that \ms1455 ~ exhibits one lensed arc inside the northern edge, 
at $20 \arcsec$ from the cluster central galaxy 
(see e.g. Le Fevre et al. 1994; Luppino et al. 1999)
(red diamond in Fig.~\ref{fig:optical}).
Besides the strong surface brightness edge to the north,
the isointensity contours show that the clusters is 
elongated to the South-West.
Moreover, in a sector from 
$\approx  150$\degd~ to $\approx  230$\degd  ~(hereafter southern-sector), 
the surface brightness shows a
 smaller surface brightness edge at   
$r\approx 90^\second$ ($450~h_{50}^{-1}$~kpc) from the X-ray peak.

\begin{inlinefigure}
%\begin{figure}
\centerline{\includegraphics[width=0.95\linewidth]{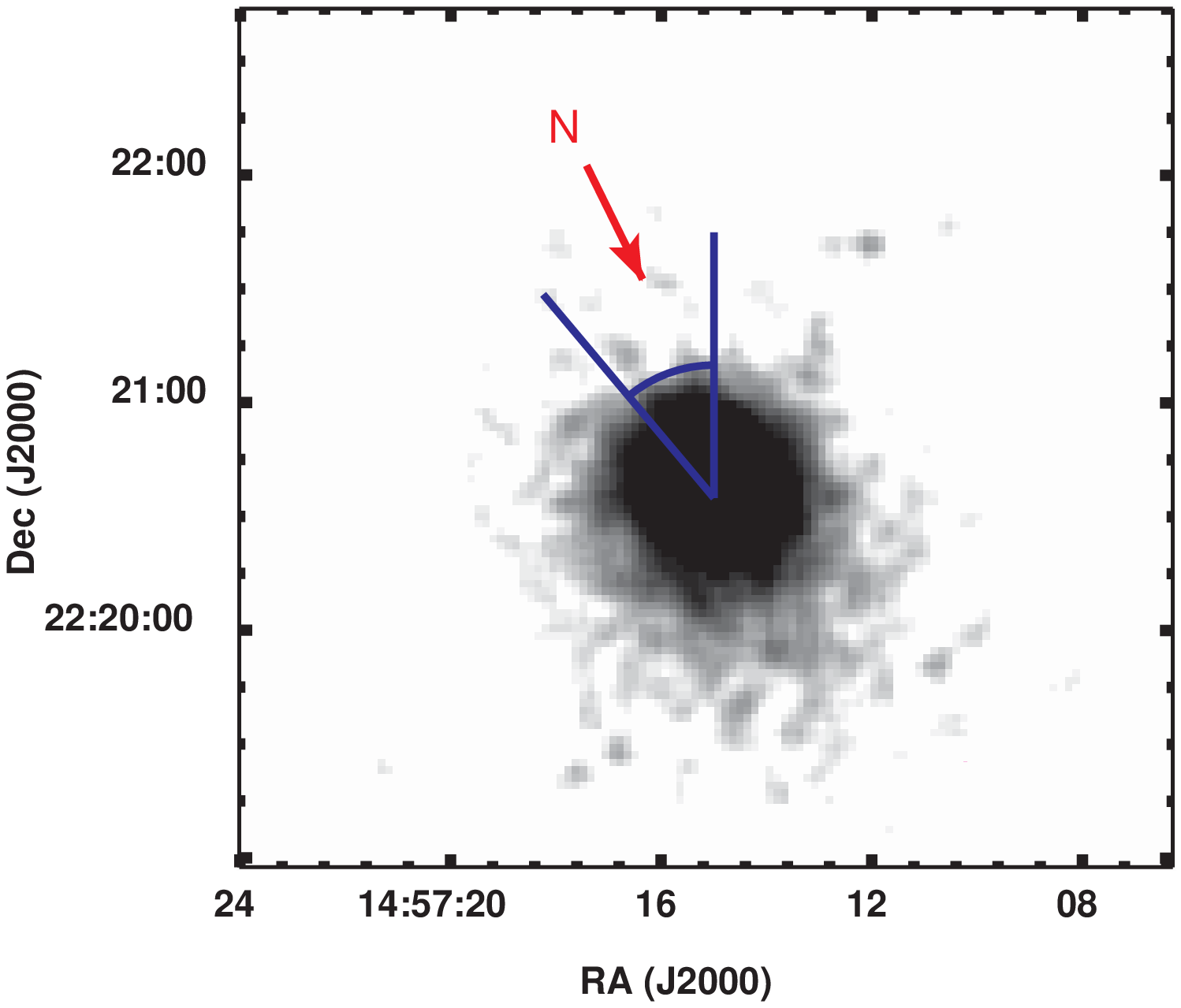}}
%\plotone{f2.eps}
\caption{Photon image in the
0.5-2.5~keV energy  band, binned to $2\arcsec$ pixels, after Gaussian 
smoothing with $\sigma=1$~pixel. 
 The arrow indicate the northern surface brightness edge.}
\label{fig:photon}
%\end{figure}
\end{inlinefigure}

\subsubsection{Density Profile}\label{par:profile}

To study the gas  density distribution across  the 
surface brightness edges, we extracted the surface brightness profiles,
in the $0.5-8$~keV energy band,
from the  north and south
sectors defined above and shown  
in Fig.~\ref{fig:optical}. From each sector 
we extracted the surface brightness profile 
using annuli centered in the X-ray peak.  
The  surface brightness profiles of the northern and southern sectors
are shown in  Fig.~\ref{fig:nprofile}a and
Fig.~\ref{fig:sprofile}a, respectively.
From these figures it is possible to see that the northern and the
southern edges correspond to 
surface brightness jumps
of factor $\approx 10$ and $\approx 3$, respectively.
From the same figures, it is clear that the first derivative of the
surface brightness profile is discontinuous on angular scales $<10\arcsec$.
This particular shape of the brightness profiles may indicate  
a discontinuity in the gas density profile. 

%\begin{figure}
\begin{inlinefigure}
\centerline{\includegraphics[width=0.95\linewidth]{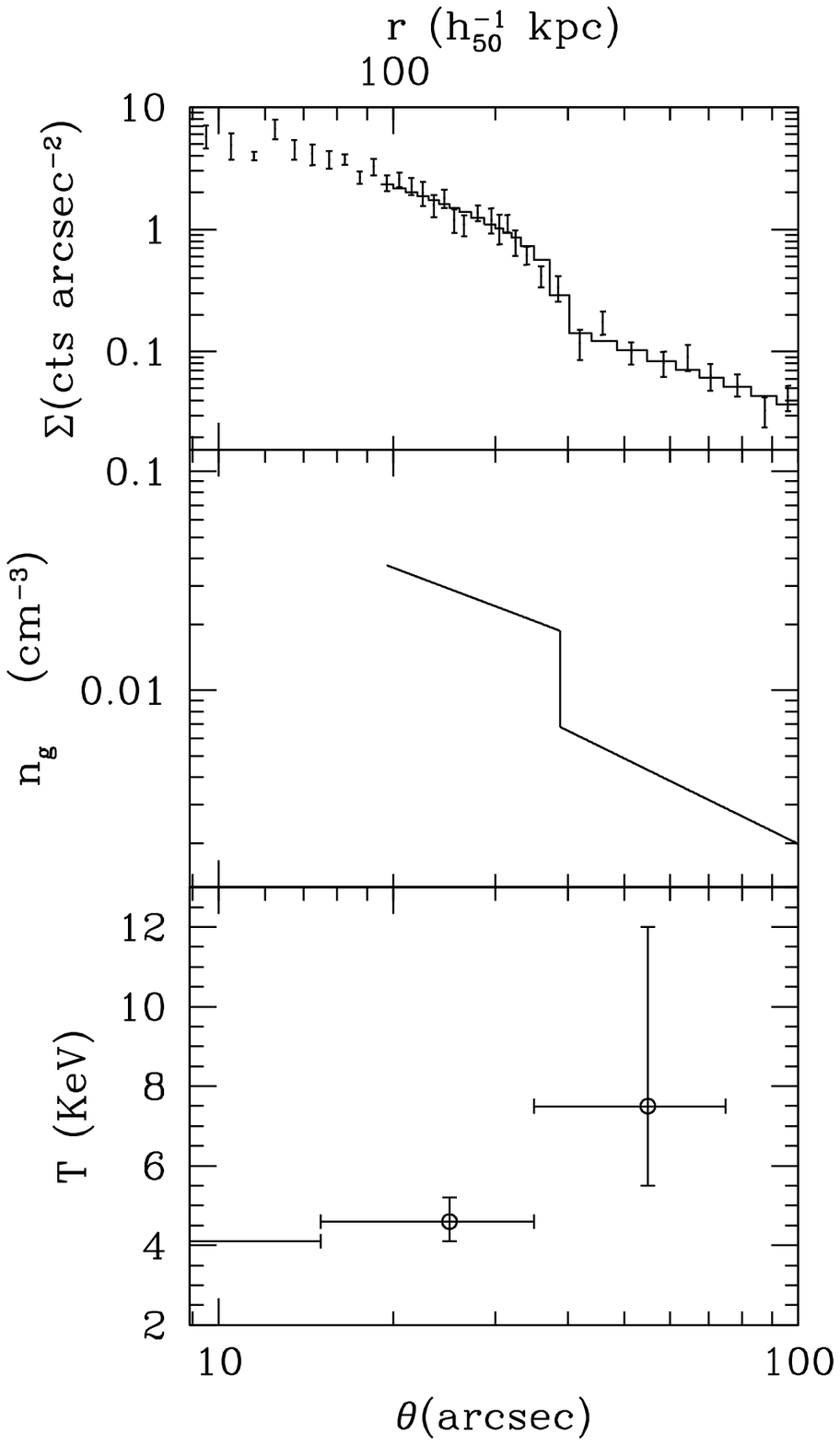}}
%\plotone{f3.eps}
\caption{X-ray surface brightness, gas density model, and temperature  
profile across the density jump in the north sector. 
{\it a)} -- X-ray surface brightness 
profile across the northern density jump. 
 Errors are $1\sigma$. The histogram
is the best fit brightness model that corresponds to the gas density
model shown in the middle panel. {\it b)} --
 The best fit gas density model for the X-ray
surface brightness
profile of the upper panel. {\it c)} --
Temperature profile obtained from spectra extracted in a sector from
  -15\degd~ to 60\degd . Error bars are at $68\%$ confidence level.}
\label{fig:nprofile}
\end{inlinefigure}
%\end{figure}

To quantify this discontinuity, we fit both brightness
profiles
with a simple radial density model composed of  two 
power-laws:
\begin{equation}
n_0\propto \left\{
\begin{array}{*{2}{l}}
 A_{jump}\left ({r / r_{jump}} \right)^{-\alpha}; & r < r_{jump};  \\
&\\
 \left ({r / r_{jump}} \right)^{-\alpha_1}; & r \ge r_{jump};  \\
\end{array} \right . 
\end{equation}
The 
model is characterized  by a density discontinuity (``jump'') 
by a factor $A_{jump}$
at the radius $r_{jump}$.  
For simplicity, we projected the density model
under the assumption of spherical symmetry. 
We ignore any gas temperature variation which
represent a small correction ($\le 0.5\%$) for the energy 
band  we are using. 
The free parameters are the power law slopes  and
the position and the  amplitude of the jump. 
To perform the fit we used the Sherpa
package (A. Siemiginowska, in preparation).
The best fit
values with their $90\%$ errors are reported in Table 1. 
The best fit  density
models are shown in Fig.~\ref{fig:nprofile}b and
Fig.~\ref{fig:sprofile}b 
 while the 
corresponding brightness profiles
are overlaid as histograms on the data points on
Fig.~\ref{fig:nprofile}a and Fig.~\ref{fig:sprofile}a.
We find that the best fit 
density  jump factors 
are $2.7$ and $1.7$ for the northern and the southern 
density profiles, respectively.

%%%%%%%%%%%%%%%%%%%%%%%%
% Table 1
%%%%%%%%%%%%%%%%%%%%%%%%

\bigskip

{\footnotesize
{\centerline {\bf TABLE 1}}
{\centerline {Density Model Fit}}
\noindent
\begin{tabular}{ c c c c c c }
\hline \hline

Sector  & $\alpha$ & $\alpha_1$ & $A_{jump}$ & $r_{jump}$ &
$\chi^2/d.o.f.$ \\
 & &  & & (arcsec) & \\
\hline 
\\
N & $1.00\pm 0.08$ & $1.3\pm 0.1$ & $2.7\pm 0.1 $  & $38\pm 2$ & 17.5/19 \\
S & $1.1\pm 0.2$   & $1.6\pm 0.2$ & $1.7\pm 0.1 $  & $93\pm 3$ & 3.5/5 \\
%  &$ \pm 0.08$&  $\pm 0.1$ & $\pm .1 $& $ \pm 2$ & ??? \\
\\
\hline
\end{tabular}
}
\bigskip

%\begin{figure}
%\plotone{f4.eps}
\begin{inlinefigure}
\centerline{\includegraphics[width=0.95\linewidth]{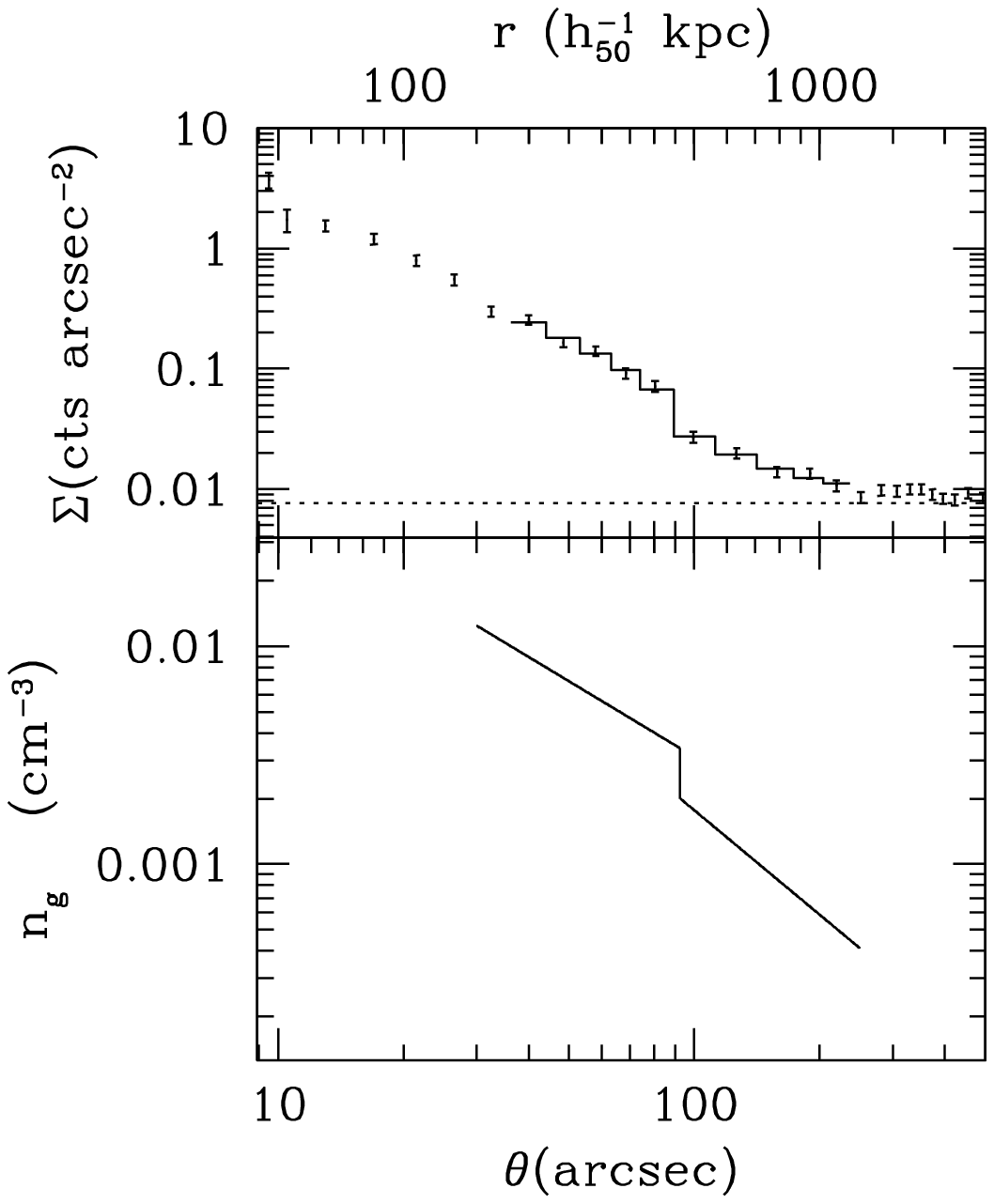}}
\caption{X-ray surface brightness and gas density model
across the density jump in the south sector. 
{\it a)} -- X-ray surface brightness 
profile across the southern density jump. 
 Errors are $1\sigma$. The histogram
is the best fit brightness model that corresponds to the gas density
model shown in the b) panel. 
The horizontal dotted line indicates the background level.
{\it b)} --  The best fit gas density model for the X-ray
surface brightness profile of the panel a.}
\label{fig:sprofile}
\end{inlinefigure}
%\end{figure}

%
%
%
\subsection{Spectral Analysis}\label{par:spectral}

Spectra were extracted in the 0.8-9.0~keV band in PI channels 
that correct for the gain difference
between the different regions of the CCDs. The spectra were then
grouped to have a minimum of 50 net counts per bin and fitted using the
XSPEC package (Arnaud 1996). To extract spectra we used the 
 CIAO 2.1.1 package (M. Elvis, in preparation).
Both the redistribution matrix (RMF) and the
effective area file (ARF) for all the CCDs are  position dependent. In our
spectral analysis we computed  position dependent 
RMFs and ARFs and  weighted them
by the X-ray brightness over the corresponding image region 
using ``calcrmf'' and ``calcarf'' software \footnote{A. Vikhlinin 2000, 
(http://asc.harvard.edu/ ``Software Exchange'', ``Contributed Software'').}.

%\bigskip
{\footnotesize
\begin{table*}
{\centerline {\bf TABLE 2}}
{\centerline {Comparison of Spectral Fits for the innermost 200\arcsec region}}
\noindent
\begin{tabular}{l c c c c c}
\hline \hline
\\
Model  &  $kT$ (keV) & $n_H$ ($10^{20}$~cm$^{-2}$) & $Z$ (Solar) &
$\dot M$ ($M_\odot$~yr$^{-2}$) & $\chi^2/$d.o.f. \\
%       &  (keV)& ($10^{20}$~cm$^{-2}$)& (Solar) & ($M_\odot$~yr$^{-1}$) & \\
\hline 
\\
WABS$\times$MEKAL & $4.8^{+0.4}_{-0.3}$ & $3.1$ & $0.28^{+0.1}_{-0.1}$ & --
& 105/103 \\
\\
WABS$\times$MEKAL & $5.0^{+0.6}_{-0.6}$ & $<5.6$ & $0.28^{+0.1}_{-0.1}$ & --
& 105/102 \\
\\
WABS$\times$(MEKAL & $5.0^{+1.0}_{-0.5}$ & $3.1$ & $0.28^{+0.1}_{-0.1}$ & $<908$
& 105/102 \\
+MKCFLOW) & & & & & \\
\hline
\\
\end{tabular}
\\
\end{table*}
}
%\bigskip

To account for the flux discrepancy between the back-illuminated and
front-illuminated chips that characterize the current (as of June 2001)
combination of spectral response matrices and quantum efficiency
curves (see Vikhlinin 2000) we multiplied all the ARFs by a
position independent fudge factor of 0.93 for $E<1.8$~keV.

%\begin{figure}
\begin{inlinefigure}
\centerline{\includegraphics[width=0.95\linewidth]{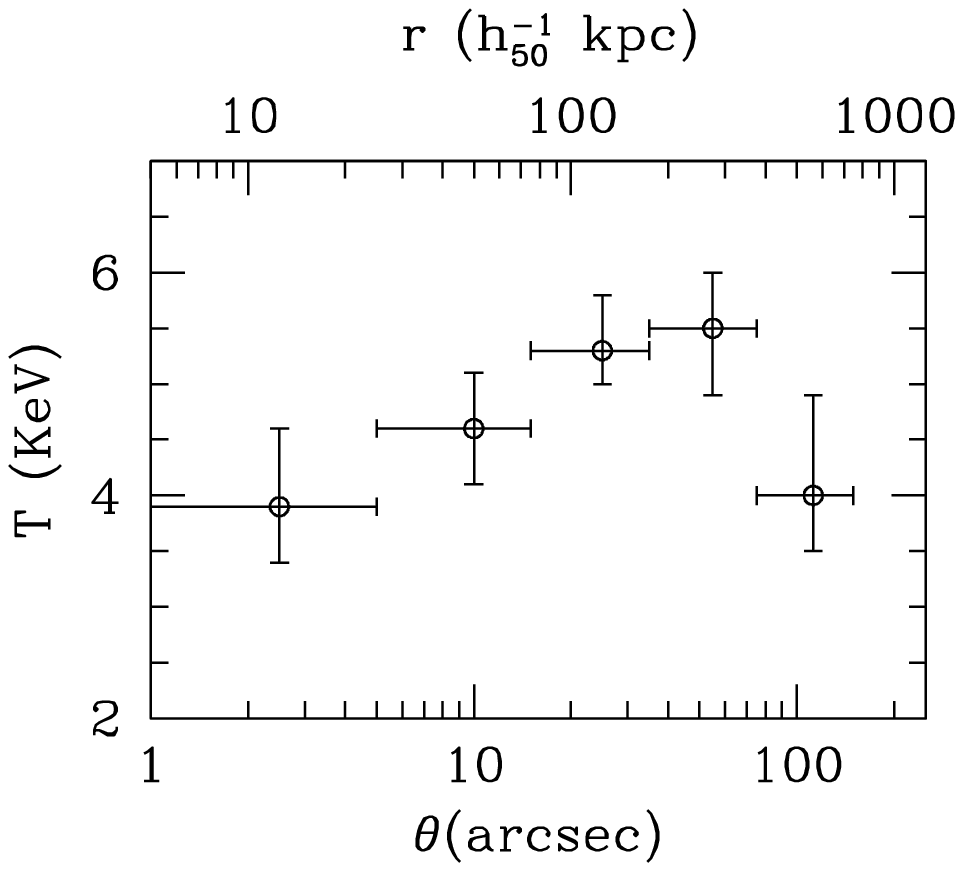}}
%\plotone{f5.eps}
\caption{Cluster temperature profile in circular annuli
centered on the X-ray peak. For each annulus we fitted the spectra
using an absorbed single temperature model with $n_h$ and metallicity 
fixed at the galactic value and  0.27 solar, respectively. 
Error bars are at $68\%$ confidence level.}
\label{fig:temperature}
\end{inlinefigure}
%\end{figure}

\subsubsection{Average Temperature}\label{par:average}

To check for consistency with previous observations, we 
 extracted an overall spectrum for  \ms1455 ~ in a circular region
 with $r=200\arcsec$ centered on the X-ray peak
 and  fitted it using different models as listed in Table~2.
First, we used  an absorbed 
 single temperature thermal Mekal model (see e.g. Kaastra, 1992; 
Liedahl, Osterheld, \& Goldstein, 1995; and references therein) fixing
 the  equivalent hydrogen column  to the Galactic value 
($n_H=3.1\times 10^{20}$~cm$^{-2}$). We find that the resulting
 temperature $T=4.8_{-0.3}^{+0.4}$~keV is  within  the error of the 
temperature 
obtained from ASCA ($T=5.01_{-0.26}^{+0.26}$, see Allen et al.
 1996). Freeing $n_H$ 
 does not  improve significantly  the spectral fit; the best fit
 $n_H$ is consistent with the galactic
 value and is significantly lower than the ASCA result
 $n_H=(9.0\pm1.7)\times 10^{20}$cm$^{-2}$ (Allen et al. 1996).
Finally, we added a cooling flow component. In this case the
 abundance of heavy elements is linked between the two components and
 the lowest gas temperature is fixed at 0.02~keV.  
We find no significant improvement for  $\chi^2$ with respect to the 
single temperature model. Because we restricted our spectral analysis 
to $E>0.8$~keV, our data do not allow us to constrain very well the
cooling flow mass deposition rate, but we can place a $90\%$ 
upper limit of $910\dot M_\odot$~yr$^{-1}$,
 significantly lower than the cooling 
rate inferred from the \asca ~ data 
($2040^{+720}_{-880}\dot M_\odot$~yr$^{-1}$; Allen et al. 1996). 
We notice that this result appears to be consistent with the finding
of other authors who show that the cooling rates derived from 
\chandra ~ data using the standard cooling flow model 
are 5-10 times lower than earlier
estimates (see e.g. McNamara et al. 2000; David et al. 2000; but see
Allen et al. 2000).

The total absorbed cluster flux in the  2-10~keV  energy band, 
measured with  ACIS, is $f_X=4.0 \times
10^{-12}$~erg~cm$^{-2}$~s$^{-1}$, 
which  is 8\% higher than the \asca ~ value (Allen et al. 1996).
 Conversely the flux in the 0.1-2.4~keV energy band
is  $f_X=4.8 \times
10^{-12}$~erg~cm$^{-2}$~s$^{-1}$ consistent with the \rosat ~ 
value (Ebeling et al. 1998).  
The flux corresponds to a 
luminosity at 0.1-2.4~keV in the source rest frame of 
$L_X=2.0\times 10^{45}$ erg~s$^{-1}$. 

%Finally, we notice that from the
%$L_X-T$ relation in the 0.1-2.4~keV band 
%(see e.g., Markevitch 1998), we find that the
%luminosity of \ms1455 ~ is typical of a 
% $T\approx (9.5\pm1.5)$~keV cluster, a temperature much higher than 
%measured.

\subsubsection{Temperature Profile}\label{par:tprofile}

To measure the temperature profile we 
divided the cluster into five circular annuli centered in the X-ray
peak. We extracted the spectra from each annulus and we fitted them
with an absorbed  single temperature  Mekal model.
We fixed the  equivalent hydrogen column  to the Galactic value 
and the metallicity to the cluster mean $Z=0.28 Z_\odot$ .
The resulting temperature profile is shown in Fig.~\ref{fig:temperature}. 
The cluster appears to be non-isothermal as it  shows a
temperature decrement  both in the cluster core ($r<100h_{50}^{-1}$~kpc) and
perhaps  at $r>500h_{50}^{-1}$~kpc.

To understand the nature of the observed surface brightness edges we
need to measure the temperature variation across the edges.
Because of the short exposure of this observation, 
the statistics are rather poor.
To measure the temperature profile in the northern sector we
extracted  spectra in a slightly  larger sector whose
angles are from   -15\degd~ to 60\degd.
In Fig.~\ref{fig:nprofile} we report the result of our fits.
%Even though we increased the angular dimension of sector,
%the statistical errors are still rather too large to draw any 
%definitive conclusion on the temperature jump. 
%However, we  speculate that 
The profile suggests a temperature 
rise as one moves outward across the northern brightness edge,
although with marginal significance.
%Because the southern edge is located at larger radii than the northern edge,
The present observation  yields no 
statistically significant temperature difference 
across the southern edge.

\section{Discussion}\label{par:discussion}

\subsection{Nature of the surface brightness edges}\label{par:nature}

In $\S$~\ref{par:image} we noted that the X-ray image of \ms1455 ~  presents
two surface brightness sharp edges one to the north and another to
the south of the X-ray peak. 
These edges correspond to  surface brightness jumps of a factor
$\approx 10$ and $\approx 3$ for the northern and southern edges, 
respectively.
In $\S$~\ref{par:profile} we argue that
these edges are consistent with jumps
in the cluster density profile of factors $2.7$ and $1.7$, respectively.
Such  density discontinuities can either indicate the presence of 
shock fronts in a gas flow (see Landau \& Lifshitz 1959) 
or the presence of cold fronts like those observed in 
A2142, A3667, RX J1720.1+2638, and A2256 (Markevitch et al. 2000;
Vikhlinin et al. 2000a, Mazzotta et al. 2001, Sun et al. 2001).

To distinguish among the two phenomena it would be sufficient to measure 
the gas temperature on   opposite sides of the gas discontinuities. 
In a shock front the temperature on the 
external side of the edge  is lower than the temperature on the
internal  side, in a cold
front the temperature variation goes in the opposite direction.

From the present data it is not possible to 
identify the nature of the southern edge,
due to the short exposure.
% there are not enough photons close to the
%southern edge to obtain statistically significant temperature
%measurements. 
%Further long exposure observations are needed 
% to disentangle the cold front or shock front hypothesis.

%Unlike the southern edge,
In  $\S$~\ref{par:tprofile} we argued that, even though the 
statistics are rather poor, the temperature profile of the northern sector
 indicates a temperature rise as one moves outward across the edge.  
By using the temperature and density
measurements across the edge, it  appears  unlikely that the observed
northern edge may be a shock front.
In fact, if we apply the Rankine-Hugoniot shock jump condition (see Landau \&
Lifshitz 1959,\S\S 82-85), to the factor of $\approx 2.7$ density jump with
the post shock temperature of $4.5$~keV  (the inner region of
the SE edge), we would expect to find a $T\approx 1.6$~keV gas in
front of the shock. This temperature is lower than that observed at $2.5
\sigma$ confidence level (see Fig.~\ref{fig:nprofile}c). 
We conclude that the northern edge is most
likely  a cold front as has been observed in other clusters by \chandra .
%Below we discuss the implication of the presence of the observed cold
%front in this cluster of galaxies.

\subsection{Cold front and cluster dynamic}\label{par:dynamic}

The presence of a cold front to the north of the cluster may  indicate that 
\ms1455 ~ hosts a moving group size ($r\approx 38\arcsec \approx 190$~kpc) 
gas cloud. Because the observed edge is
particularly sharp, we may expect that the gas cloud direction of motion
is close to the plane of sky  (Mazzotta et al. 2001).  
Vikhlinin et al. (2000a) showed that  the gas cloud  speed 
 depends 
only on the pressure variation across the 
edge.  Unfortunately, the errors on the 
temperature estimates are too large to 
obtain a statistically interesting velocity measurement.
However, in $\S$\ref{par:image} we noted that the cluster is quite
 azimuthally symmetric and that the
X-ray peak is coincident with the central cD galaxy.
These 
properties are similar to those found in
RX J1720+2638 (Mazzotta et al. 2001).
Thus, because  the gas cloud is moving in a plane close to the
plane of the sky, it is likely that, as for RX J1720+2638, 
it is slowly oscillating around the 
minimum of the potential well. 
If the gas in the cloud is associated with a similar-sized dark
matter subclump, then we may assume that the gas
approximately traces  the underlying 
dark-matter halo. 
This means that the cluster may host a moving 
group sized subclump in its center.

One possible explanation for the presence of this central  moving subclump 
is that the cluster experienced a recent  merger.
In this scenario the moving  subclump is the remaining part
of the merging object that  already has passed through the cluster 
center several times.
This may indicate that the subclump has a quite compact structure 
otherwise it would be easily destroyed by the tidal forces during multiple
core crossings.
From the deprojection analysis of Allen et al. (1996) we see that
the integrated gravitational mass at $190~h_{50}^{-1}$~kpc
is $\approx 1/3$ of the gravitational mass at $1000~h_{50}^{-1}$~kpc.
If we assume that most of the gravitational mass inside 
$190~h_{50}^{-1}$~kpc belongs to the merger subclump, than we can
estimate that the ratio between the masses of the merging objects was
something between 3 to 5. Moreover, as 
the moving subclump shows no offset with respect to the cluster 
centroid, we may also assume that it was a head-on
merger.
In $\S$~\ref{par:average} we show that, even though the inferred mass
deposition rate obtained from the \chandra ~ data is significantly
lower than that obtained from \asca , we cannot exclude that
this cluster hosts a massive (or moderate) cooling flow. 
Regardless of the presence of a cooling flow, it  contains a
distinct cool gas cloud.
If  \ms1455 ~ is a merger, 
then  it is not clear if that a massive cooling flow (or cool central
gas) would have survived the merger (see e.g. Roettiger, Loken,
\& Burns 1997, but see also Fabian \& Daines 1991).

%to form such a massive flow the model require that   
%the cluster  experienced no energetic gas
%perturbation in the last $10^8- 10^{9}$~yr.

An alternative scenario to explain  the presence of the
moving subclump is that, as proposed for RX J1720+2638, 
\ms1455 ~  is the 
result of the collapse of two different  perturbations 
in the primordial density field
on two different linear scales at nearly the same location in space.
As the density field evolves, both perturbations start to collapse. 
 The small scale perturbation collapses first 
and forms a central group of galaxies while the larger perturbation
continues to evolve to form 
a more extended
 cluster potential.
The central group of galaxies could have formed  slightly offset
from the center of the cluster and is now
falling into  or oscillating around the
minimum of the cluster potential well. This motion is responsible for
the observed surface brightness discontinuity.
As the initial position of the subclump lies well within  
the  main cluster, we may also  expect the velocity of the infalling 
subclump to be  subsonic.
In the hierarchical 
cosmological model for structure formation,
density perturbations on different linear scales evolve independently
and on different time scales. 
So, at the time when the group had already formed,
the larger scale perturbation was still
expanding with the Hubble flow.
%, hence, the central group 
%may have only weakly affected by its environment.
In this scenario, any cooling
flow may have started to develop  in the collapsed  
group of galaxies much before the the cluster virialized.
Because of the low speed of the central group,
 the cooling flow would not have been
destroyed by  infall into the cluster center but rather might have been
amplified by the gas compression induced by the infall of the 
surrounding mass
of the main cluster.

Regardless of the cluster formation history it appears clear that the
surface brightness edge strongly indicate
a dual gravitational  potential structure for this system.
This would be consistent with  a recent finding by 
Williams \& Navarro (1999) who claim that the
giant arc observed in \ms1455 ~(and in other clusters too)
can  be explained in terms of cold dark matter halos only if the mass
at the center is significantly increased by the presence of a massive
central galaxy (or small group).

\subsection{Cold fronts and X-ray mass estimates}\label{par:mass}  

%We  conclude this paragraph with a short remark
%about the discrepancy between X-ray and strong lensing mass estimate.

While there is  good agreement between
the  X-ray and the weak lensing mass estimates, for some systems 
the X-ray mass estimate is a factor $\approx 2$ lower than that  
obtained from strong lensing (see e.g. Miralda-Escud\'e, \&  Babul 1995, 
but see Allen 1998; and reference therein).
\ms1455 ~ is  one of these peculiar systems.
The mass estimate based on the arc in the northern sector 
(see Fig.~\ref{fig:optical}) 
is a factor of 1.8 times larger
than the X-ray estimate obtained from \asca ~ and \rosat ~ 
observations (Wu \& Fang 1997; Wu 2000).

The arc on which the strong lensing measure is
based is just inside the northern edge. This is particularly
interesting as Mazzotta et al. (2001) pointed out
that, if a cold front is present in a cluster of galaxies,
the X-ray mass profile could be underestimated at radii smaller than 
the position of the density discontinuity.
The X-ray mass underestimate depends strongly on the 
temperature variation along the edge. 
Unfortunately our temperature measurements are not accurate enough to
estimate this effect. A longer exposure observation is needed to 
verify  if this effect may account for the observed mass 
discrepancy.

\section{Conclusion}\label{par:conclusion}

We have presented the results of a  short \chandra ~ observation of the 
cluster of galaxies \ms1455 . The data show two surface
brightness edges on opposite side of the X-ray peak consistent with  
discontinuities in the density profile.
The structure of the edges is  similar to the ``cold fronts''  observed
by \chandra ~ in the  clusters A2142, A3667, RX J1720.1+2638, and A2256. 
Even though the low exposure of this observation
limits our ability to constrain the temperature jumps across the edges,
we show that the northern edge is likely to be a cold front
 produced by the subsonic motion
from the south to the north of the central group-size cloud of gas
within the cluster.

We discussed the possibility that \ms1455 ~ is an ongoing merger
with merging object mass ratio of
3-5. Moreover, the merger is most likely head-on. 
%In such a scenario
%it is not clear if the cooling flow could have survived the merger.
  
We also discussed the possibility that,  as proposed for  RX J1720.1+2638,
\ms1455 ~ is the result of the collapse of a
group of galaxies  followed by the 
collapse of a much larger, cluster-scale  perturbation at nearly the same
 location in space.
%Unlike the merger hypothesis, this last scenario would eventually explain the
%presence of a strong cooling flow.

We also mentioned that,  because of the motion of the central gas cloud,
the hydrostatic equilibrium equation may underestimate
the true cluster mass in the cluster core.
%Due to the low exposure of this observation 
%it was not possible to estimate the importance
%of this effect for this cluster.
Thus, such motion may be responsible for
the discrepancy between the X-ray and the strong lensing
mass determinations found for this system.

\acknowledgments
%We thank ............
 
P.M. acknowledges an ESA fellowship and thanks the Center for
Astrophysics for its hospitality. 
Support for this study was provided
by NASA contract NAS8-39073, grant NAG5-3064, and by the Smithsonian
Institution.

\newpage


\begin{references}

\reference{} Allen, S.\ W.\ et al.\ 1992, \mnras, 259, 67 

\reference{} Allen, S. W.,  1998, MNRAS, 296, 392


\reference{} Allen, S.\ W., Fabian, A.\ C., Edge, A.\ C., Bautz, M.\ W., 
Furuzawa, A., \& Tawara, Y.\ 1996, \mnras, 283, 263 

\reference{}  Allen, S.\ W.,  Taylor, G.B.,  Nulsen, P.E.J.,
 Johnstone,  R.M.,  David, L.P.,  Ettori, S., Fabian,  A.C.,  Forman, W.,
  Jones, C.,  McNamara, B.  2000, \mnras, in press (astro-ph/0101162)

\reference{} Arnaud, K.\ A.\ 1996, ASP 
Conf\ Ser.\ 101, Astronomical Data Analysis Software and Systems V, 5, 17 

\reference{} David, L.P., Nulsen, P.E.J., McNamara,  B.R.,  Forman, W., Jones,
C.,  Ponman, T.,  Robertson, B., Wise, M. 2000, ApJ, 
in press (astro-ph/0010224)


\reference{} Ebeling, H., Edge, 
A.\ C., Bohringer, H., Allen, S.\ W., Crawford, C.\ S., Fabian, A.\ C., 
Voges, W., \& Huchra, J.\ P.\ 1998, \mnras, 301, 881 

\reference{} Fabian, A.C, \& Daines, S.J. 1991, \mnras, 252, 17

\reference{}    Kaastra, J.S. 1992, An X-Ray Spectral Code for
Optically  Thin Plasmas (Internal SRON-Leiden Report, updated version 2.0)


\reference{} Landau, L.D., \& Lifshitz, E. M. 1959, Fluid Mechanics
(London: Pergamon)

\reference{} Le Fevre, O., Hammer, F., Angonin, M.~C., Gioia, I.~M., \& 
Luppino, G.~A.\ 1994, \apjl, 422, L5 

\reference{}  Liedahl, D.A., Osterheld, A.L., and Goldstein, W.H. 1995, ApJL, 438, 115



\reference{} Luppino, G.~A., Gioia, 
I.~M., Hammer, F., Le F{\` e}vre, O., \& Annis, J.~A.\ 1999, \aaps, 136, 117 


%\reference{} Markevitch, M.\ 1998, \apj, 504, 27 


\reference{} Markevich, M. et al. 2000a, CXC memo
(http://asc.harvard.edu/cal ``ACIS'', ``ACIS Background'')

\reference{} Markevitch, M., Ponman, T. J., Nulsen, P. E. J., Bautz,
M. W.,  Burke, D. J., David, L. P., Davis, D.,  Donnelly, R. H., 
    Forman, W. R., Jones, C., Kaastra, J.,  Kellogg, E.,  Kim, D.-W.,
Kolodziejczak, J.,  Mazzotta, P., Pagliaro, A.,  Patel, S.,
VanSpeybroeck, L.,
    Vikhlinin, A., Vrtilek, J.,  Wise, M.,  Zhao, P. 2000b, ApJ, 541, 542 


\reference{}  Mazzotta, P.,   Markevitch, M.,
 Vikhlinin, A.,  Forman, W. R.,   David, L. P., \&
 VanSpeybroeck, L., 2001a, ApJ, 555, 205

\reference{} McNamara, B.\ R.\ et al.\ 2000, \apjl, 534, L135 

\reference{} Miralda-Escud\'e, J.~\& Babul, A.\ 1995, \apj, 449, 18 

\reference{} Roettiger, K.,  Loken, C. \& Burns J.O.  1997, \apj, 109,
307

\reference{} Sarazin, C. L. 1988, X-Ray Emission in Cluster of
Galaxies (Cambridge: Cambridge Univ. Press)

\reference{} Sun, M.,  Murray, S.S.,  Markevitch, M., Vikhlinin, A., 2001, ApJ,
 submitted (astro-ph/0103103)

\reference{} Vikhlinin, A., Markevitch, M., Murray, S.S. 2000a, ApJ, 551, 160

\reference{} Vikhlinin, A., Markevitch, M., Murray, S.S. 2000b, \apjl, 549, L47

\reference{} Williams, L.~L.~R., Navarro, J.~F., \& Bartelmann, M.\ 1999, 
\apj, 527, 535 

\reference{} Wu, X.~\& Fang, L.\ 1997, \apj, 483, 62 

\reference{} Wu, X.\ 2000, \mnras, 316, 299 


\end{references}
\end{document}